\documentclass[%
 reprint,
 amsmath,amssymb,
 aps,
]{revtex4-2}

\usepackage{graphicx}
\usepackage{hyperref}
\usepackage{dcolumn}
\usepackage{bm}
\usepackage{siunitx,physics,multirow,bigdelim}

\begin{document}
\newcommand{\HEO}{Mg$_{0.2}$Co$_{0.2}$Ni$_{0.2}$Cu$_{0.2}$Zn$_{0.2}$O}
\newcommand{\etal}{\textit{et al.}}
\newcommand{\ie}{\textit{i.e}.}
\newcommand{\eg}{\textit{e.g}.}
\newcommand{\etc}{\textit{etc}.}
\preprint{APS/123-QED}

\title{Localization of vibrational modes in high-entropy oxides}

\author{C.~M.~Wilson}
\email{cw14mi@brocku.ca}
\affiliation{Department of Physics, Brock University, St. Catharines, Ontario L2S 3A1, Canada}
\author{R.~Ganesh}%
\affiliation{Department of Physics, Brock University, St. Catharines, Ontario L2S 3A1, Canada}
\author{D.~A.~Crandles}
\affiliation{Department of Physics, Brock University, St. Catharines, Ontario L2S 3A1, Canada}
\date{\today}

\begin{abstract}
The recently-discovered high-entropy oxides offer a paradoxical combination of crystalline arrangement and high disorder. They differ qualitatively from established paradigms for disordered solids such as glasses and alloys. In these latter systems, it is well known that disorder induces localized vibrational excitations. In this article, we explore the possibility of disorder-induced localization in \HEO, the prototypical high-entropy oxide with rock-salt structure. 
To describe phononic excitations, we model the interatomic potentials for the cation-oxygen interactions by fitting to the physical properties of the parent binary oxides. We validate our model against the experimentally determined crystal structure, bond lengths, and optical conductivity.
The resulting phonon spectrum shows wave-like propagating modes at low energies and localized modes at high energies. Localization is reflected in signatures such as participation ratio and correlation amplitude. Finally, we explore the possibility of increased mass disorder in the oxygen sublattice. Admixing sulphur or tellurium atoms with oxygen enhances localization. It even leads to localized modes in the middle of the spectrum. Our results suggest that high-entropy oxides are a promising platform to study Anderson localization of phonons. 
\end{abstract}

\maketitle


\section{\label{sec:intro}Introduction}
\begin{figure}[htbp]
    \centering
    \includegraphics[width=0.45\textwidth]{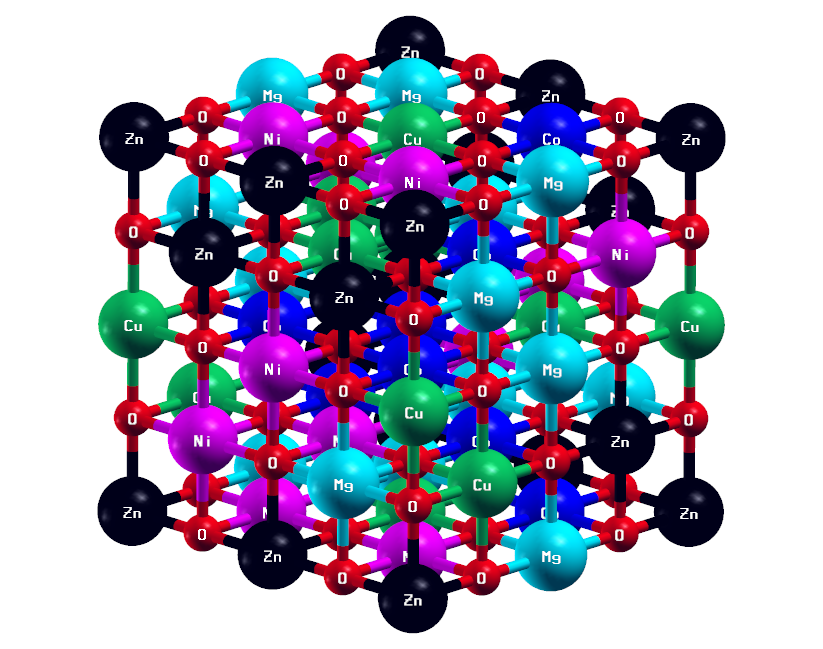}
    \caption{Possible disorder realization of the \HEO~crystal structure.}
    \label{fig:supercell}
\end{figure}

The field of high-entropy oxides began with the synthesis of \HEO~(hereafter referred to as HEO) by C.~M.~Rost~\etal~in 2015~\cite{rost_Nature_2015}. The five parent binary oxides have limited mutual solid solubility and even differ in their crystal structures. Nevertheless, Rost~\etal~showed that HEO crystallizes in the rocksalt ($F\bar{m}3m$) structure~\cite{rost_Nature_2015} with a lattice constant~$a=4.236(1)~\si{\angstrom}$~\cite{crandles_JAP_2021}. The oxygen sublattice is ordered, while the cation sublattice is uniformly occupied by Mg, Co, Ni, Cu, and Zn with no observable short-range ordering at an atomic resolution of~1--3~$\si{\angstrom}$~\cite{chellali_ScriptaMat_2019}. A possible instance of the HEO crystal structure is shown schematically in Figure~\ref{fig:supercell}. Due to their low thermal conductivity/elastic modulus~\cite{rost_AM_2018} and high Li-ion room temperature conductivity~\cite{sarkar-RCS-2019}, high-entropy oxides are attractive candidates for thermal coatings~\cite{rost_AM_2018} and solid-state batteries~\cite{sarkar-RCS-2019}.

As disordered media do not have translational symmetry, phonon excitations cannot be associated with specific wavevectors. Likewise, there is no conventional classification of modes as longitudinal/transverse or acoustic/optical, although these notions can be suitably extended~\cite{beltukov_PRE_2016,bell_JPhysC_1975}. In the context of glasses, Allen and Feldman~\cite{allen_PRB_1993} and Allen~\etal~\cite{allen_PhilMagB_1999} proposed a new taxonomy, with phonons classified as propagons, diffusons and locons. Propagons are extended, plane-wave-like excitations which generally occupy the low-frequency part of the vibrational density-of-states (VDOS). Within a propagon mode, atomic displacements are spatially coherent; \ie~nearest-neighbours tend to vibrate in the same direction. Diffusons occur in the middle region of the VDOS, with components that are randomly-oriented. While they are spatially extended, a wavepacket of diffusons transports vibrational energy diffusively. Locons appear at high frequencies, above a `mobility-edge'~\cite{allen_PhilMagB_1999}. They exhibit randomly-oriented displacements that are localized over a small cluster of atoms.

The scaling theory of localization~\cite{abrahams_PRL_1979} predicts that all eigenstates of a one- or two-dimensional system become exponentially localized when any amount of disorder is added to an otherwise clean system. In three-dimensions, eigenstates may be localized or extended, depending upon the amount of disorder present in the system. This physics has been seen in the vibrational excitations of amorphous solids such as $\alpha$-silicon.

The disorder in amorphous systems is structural in character. This may be contrasted with mass- and spring-constant-disorder in systems such as high-entropy alloys -- a family of materials with atoms of metallic elements randomly distributed on a background lattice~\cite{yeh_AEM_2004}. However, as alloys are typically electrically conducting, the physics of phonon localization is complicated by the inevitable presence of electron-phonon coupling. High-entropy oxides offer an attractive alternative. They possess a high degree of mass and spring-constant disorder. At the same time, they are electrically insulating with a large gap to electronic excitations. This makes for a relatively clean platform to study the localization of vibrational modes.

This paper is organized as follows: in Sec.~\ref{ssec:methods}, details of the classical lattice dynamics simulations are discussed. Sec.~\ref{ssec:model} presents our model for \HEO~and discusses refinements over previous approaches. Sec.~\ref{ssec:validation} compares the simulated crystal structure, bond lengths, and optical conductivity of HEO to experiment. Sec.~\ref{ssec:characterization} characterizes the vibrational modes in terms of the relative amplitude~\cite{carvalho_PRB_2006}, phase quotient~\cite{bell_faraday_1970}, and polarization. Sec.~\ref{ssec:localized-modes} presents several diagnostics for mode localization which suggest that locons do indeed exist in HEO beyond a high-frequency mobility edge in the VDOS.

\section{\label{sec:methods}Model, validation, and methods}

\subsection{\label{ssec:methods}Methods}
We use the \texttt{General Utility Lattice Program (GULP)}~\cite{gale_MS_2003} to find the equilibrium structure and to evaluate the vibrational spectrum. This program is known to reproduce experimental structure and phonon band structure for ionic solids such as MgO and ZnO~\cite{tipaldi_LJP_2022}. It takes into account Coulomb interactions as well as short-ranged forces. Supercells of HEO containing up to 8000 atoms were generated by uniformly distributing equiatomic proportions of Mg, Co, Ni, Cu, and Zn throughout the cation sublattice. 

\subsection{\label{ssec:model}Model}
The effects of polarizability were included with the shell model introduced by Dick and Overhauser~\cite{dick_PR_1958}, wherein atoms are divided into massive point charges bonded by harmonic springs to massless, charged shells; \ie~a core-shell pair interact through a potential
\begin{equation}
    \Phi=\frac12\,kr_{cs}^2,
    \label{eq:core-shell}
\end{equation}
where $k$ is the stiffness constant and $r_{cs}$ is the core-shell distance. For simplicity, the sum of the shell charge~$Y$ and the core charge was fixed to the formal charge on each ion, either~$+2$ for cations or~$-2$ for oxygens. The short-range shell-shell interactions were parameterized by the Buckingham potential
\begin{equation}
    \Phi
    =A_{bb'}\,\exp(-\frac{\abs{b-b'}}{\rho_{bb}})
    -\frac{C_{bb'}}{\abs{b-b'}^6},
    \label{eq:buckingham}
\end{equation}
where $b,b'$ are internal atomic coordinates and the two terms on the right-hand side represent, respectively, the repulsion between electron clouds and the van-der-Waals attraction. In addition to Eqs.~\ref{eq:core-shell} and~\ref{eq:buckingham}, the Coulomb potential also acts between all pairs of cores and shells.

Lewis and Catlow~\cite{lewis_JPhysC_1985} observed reasonable agreement between the simulated and experimental elastic constants of spinel oxides by a) transferring the potential parameters directly from the parent binary oxides; b) neglecting cation-cation interactions, and; c) setting $C_{bb'}=0$ for all interactions except oxygen-oxygen~\footnote{Lewis and Catlow assumed oxygen was polarizable and treated all cations with the rigid-ion model. Hence there were no dipole-dipole interactions and $C_{bb'}=0$. The assumption here is that oxygen is larger and more polarizable than the cations in HEO and should therefore experience the largest dipole-dipole force.}. To reduce the number of free parameters, we make the same assumptions about HEO and use the Lewis and Catlow oxygen-oxygen potential without modification. \texttt{GULP} was then used to determine the remaining unknown parameters $\{Y,k,A,\rho\}$ by performing a relaxed fit to the parent binary oxides' experimental crystal structures, dielectric constants, and phonon frequencies. The final values of the parameters are listed in Tables~\ref{tab:buckingham} and~\ref{tab:SM}. Following Popov~\etal, two Buckingham potentials operating over different distance ranges were used to model Cu--O interactions~\cite{popov_JPhysC_1995}.

In Sec.~\ref{sss:participation-ratio}, we discuss two hypothetical materials: `high-entropy sulfide oxide' (HESO) and `high-entropy telluride oxide' (HETeO). They are generated by randomly substituting half of the oxygen sublattice in HEO with sulfur and tellurium ions. As a minimal model for HESO, we take the cation-sulfur potential to be identical to the cation-oxygen potential. Likewise, we take the sulfur-sulfur and sulfur-oxygen potentials to be identical to the oxygen-oxygen potential. The sulfur and oxygen shell model parameters are also taken to be the same. Analogous assumptions are made for HETeO. These assumptions amount to increasing the mass disorder in HEO, but not the spring-constant disorder. HESO and HETeO have not been experimentally realized, although a high-entropy oxyfluoride has recently been synthesized~\cite{lin_JournMatSci_2020}.
\begin{table}[htbp]
\caption{\label{tab:buckingham}
Buckingham shell-shell parameters. The potentials operate over a distance range specified in the rightmost two columns.
}
\begin{ruledtabular}
\begin{tabular}{c@{\,}ccccc}
&
$A~(\si{\electronvolt})$&
$\rho~(\si{\per\angstrom})$&
$C~(\si{\electronvolt\per\angstrom^6})$&
$r_{min}~(\si{\angstrom})$&
$r_{max}~(\si{\angstrom})$\\
\colrule
O--O & 22764 & 0.149 & 27.88 & 0 & 12\\
Mg--O & 1266.7 & 0.301 & 0 & 0 & 8\\
Co--O & 1244.3 & 0.305 & 0 & 0 & 8\\
Ni--O & 1794.8 & 0.283 & 0 & 0 & 8\\
\ldelim\{{2}{*}[CuO] 
  & 2054.7 & 0.269 & 0 & 0 & 2.3\\
  & 558.23 & 0.360 & 0 & 2.3 & 8 \\
Zn--O & 571.82 & 0.353 & 0 & 0 & 8
\end{tabular}
\end{ruledtabular}
\end{table}

\begin{table}[htbp]
\caption{\label{tab:SM}
Shell model parameters.
}
\begin{ruledtabular}
\begin{tabular}{ccc}
Ion & $Y~(e)$ & $k~(\si{\electronvolt\per\angstrom\squared})$\\
\colrule
O & -2.88 & 70.52\\
Mg & 2.77 & 137.3\\
Co & 3.20 & 66.51\\
Ni & 3.68 & 95.48\\
Cu & 3.82 & 85.37\\
Zn & 2.19 & 19.19
\end{tabular}
\end{ruledtabular}
\end{table}

\subsection{\label{ssec:validation}Model validation}
We next compare results from the \texttt{GULP} model to known experimental quantities to validate our model. In the following discussion, all \texttt{GULP} results were averaged over fifty 4096-atom disorder realizations before comparing with results from literature.

\subsubsection{\label{sss:crystal-structure}Crystal structure}
The HEO lattice parameters obtained from our model are listed in Table~\ref{tab:crystal-structure}. The cell angles are all correctly equal to $90^{\circ}$ within error, and the cell lengths are in pairwise agreement. The cell lengths are slightly lower than XRD values (by $\sim 0.03~\si{\angstrom}$ or $1\%$). Our results can be compared with Anand~\etal~\cite{anand_ActaMat_2018}, a theoretical study where cation-oxygen potentials developed by Lewis and Catlow~\cite{lewis_JPhysC_1985} were applied to HEO. This study reports an even shorter cell length of~$4.16~\si{\angstrom}$. This suggests that our model, with its refinements, is closer to true HEO.
\begin{table}[htbp]
\caption{\label{tab:crystal-structure}
Crystal structure of HEO. The lattice parameter from~\cite{anand_ActaMat_2018} is taken to be twice the average cation-oxygen spacing of $2.08~\si{\angstrom}$. 
}
\begin{ruledtabular}
\begin{tabular}{cccc}
& XRD~\cite{crandles_JAP_2021}
& Anand~\etal~\cite{anand_ActaMat_2018}
& This work\\
\colrule
$\alpha~(\si{\degree})$
& 90
& --
& 90.000(2)\\
$\beta~(\si{\degree})$
& 90
& --
& 90.000(2)\\
$\gamma~(\si{\degree})$
& 90
& --
& 90.000(2)\\
$a~(\si{\angstrom})$
& 4.236(1)
& 4.16
& 4.2064(3)\\
$b~(\si{\angstrom})$
& 4.236(1)
& 4.16
& 4.2062(3)\\
$c~(\si{\angstrom})$
& 4.236(1)
& 4.16
& 4.2063(3)\\
\end{tabular}
\end{ruledtabular}
\end{table}

\subsubsection{\label{sss:bond-lengths}Bond lengths}
\begin{figure}[htbp]
    \centering
    \includegraphics{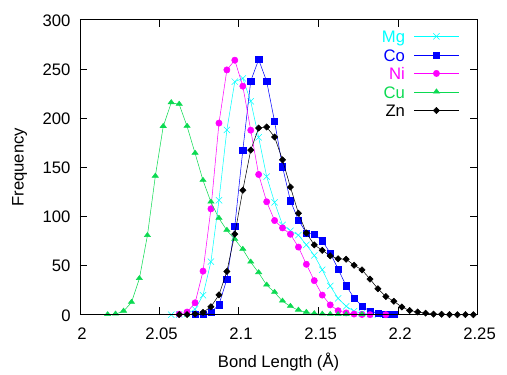}
    \caption{Distribution of nearest-neighbour cation-oxygen spacings (bond lengths) in HEO. The solid lines are guides to the eye.}
    \label{fig:bond-lengths}
\end{figure}

The distributions of nearest-neighbour cation-oxygen spacings (bond lengths) are presented in Figure~\ref{fig:bond-lengths}. All distributions are composed of a main peak and a shoulder in the right tail. This can be compared with Anand~\etal~\cite{anand_ActaMat_2018} which reports a bimodal distribution of Cu--O bonds in HEO and concluded that the experimentally-observed local lattice distortion around the Cu ions is due to a combination of the Jahn-Teller effect~\cite{berardan_JAC_2017} and differences in ionic radii. The median bond lengths are listed explicitly in Table~\ref{tab:bond-lengths} and are compared to results from EXAFS. Our results are in agreement with the experiment of Sushil~\etal~\cite{sushil_MatChemPhys_2021}.
\begin{table*}[htbp]
\caption{\label{tab:bond-lengths}
Median bond lengths in HEO. Rost~\etal~and Sushil~\etal~are EXAFS studies. Mg--O bonds were not analysed due to insufficient energy resolution. Rost~\etal~and Sushil~\etal~report two bond lengths for Cu--O, reflecting local Jahn-Teller distortions. The values cited in the table are averages, weighted by the number of bonds with each length. The error estimates from~\cite{anand_ActaMat_2018} are one-half the widths of the data points obtained by digitizing their Fig.~4a). 
}
\begin{ruledtabular}
\begin{tabular}{ccccc}
& Rost~\etal~\cite{rost_JACS_2017}
& Sushil~\etal~\cite{sushil_MatChemPhys_2021}
& Anand~\etal~\cite{anand_ActaMat_2018}
& This work\\
\colrule
Mg--O
& --
& --
& 2.088(3)
& 2.11(2)\\
Co--O
& 2.089(9)
& 2.0906
& 2.093(3)
& 2.12(2)\\
Ni--O
& 2.084(5)
& 2.0918
& 2.080(3)
& 2.10(2)\\
Cu--O
& 2.07(5)
& 2.0733
& 2.033(3)
& 2.07(2)\\
Zn--O
& 2.078(9)
& 2.0984
& 2.094(3)
& 2.12(3)
\end{tabular}
\end{ruledtabular}
\end{table*}

\subsubsection{\label{ss:sigma1}Optical conductivity}
\begin{figure}[htbp]
\includegraphics{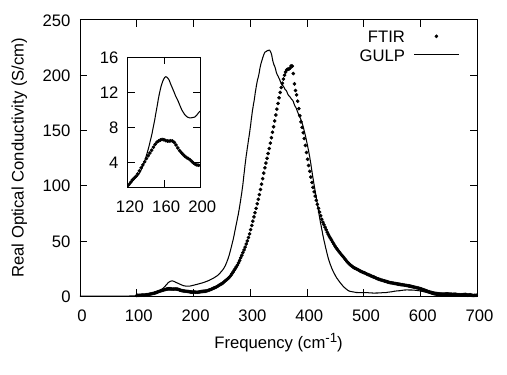}
\caption{\label{fig:optical-conductivity}Optical conductivity of HEO. Experimental FTIR (Fourier-transform infrared) data at 300~\si{\kelvin} adapted from~\cite{crandles_JAP_2021}. This data is compared against the result from \texttt{GULP} based on our model.
}
\end{figure}

Afsharvosoughi and Crandles~\cite{crandles_JAP_2021} measured the infrared reflectance of HEO and extracted the real optical conductivity $\sigma'(\omega)$ from Kramers-Kronig analysis~\cite{crandles_JAP_2021}. Their data is reproduced in Fig.~\ref{fig:optical-conductivity}. Observe that $\sigma'(\omega)$ consists of a strong mode near~$360~\si{\per\centi\meter}$ and a weak mode at $160~\si{\per\centi\meter}$. The strong mode was identified with the reststrahlen band characteristic of rocksalt crystals. The weak mode is of yet unclear origin.

Figure~\ref{fig:optical-conductivity} compares the optical conductivity from \texttt{GULP} to experiment. Our model successfully reproduces the weak mode and provides satisfactory agreement with the reststrahlen band.

\section{Results}
\subsection{\label{ssec:characterization}Mode characterization}
In this section the vibrational modes in HEO are characterized in terms of their phase quotient, relative amplitude, and eigenvector polarization. All results are shown for a single 4096-atom disorder realization.

\subsubsection{\label{sss:relative-amplitude}Relative amplitude}
Carvalho~\etal~define the relative amplitude~$A$ of ionic species $B$ as~\cite{carvalho_PRB_2006}
\begin{equation}
    A(B;\omega_s)
    =\sum_{b\in B}\sum_{\alpha}\abs{\varepsilon_{\alpha}(b;s)}^2
    \label{eq:RA}
\end{equation}
where $\varepsilon_\alpha(b;s)$ is the projection of the $s$th eigenvector onto the $b$th atom in the $\alpha$th Cartesian direction. The summation $\sum_{b\in B}$ extends over all atoms~$\{b\}$ belonging to the species~$B$. If the atomic motion in mode~$s$ is dominated by atoms in~$B$ then $A(B;\omega_s)\sim1$; conversely, if the vibrations due to species~$B$ are weak then $A(B;\omega_s)\ll1$. Note that $\sum_BA(B;\omega_s)=\sum_{b\alpha}\abs{\varepsilon_\alpha(b;s)}^2=1$ by the orthonormality of the vibrational eigenvectors~\cite{srivastava_1990}.

\begin{figure}[htbp]
    \centering
    \includegraphics{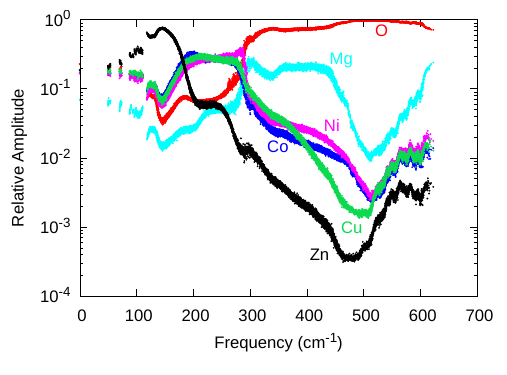}
    \caption{Relative amplitudes in HEO.}
    \label{fig:relative-amplitude}
\end{figure}

Figure~\ref{fig:relative-amplitude} shows the relative amplitudes of the six ionic species in HEO. Observe that~$A$ is large for the heavier cations at low frequency; as frequency increases the cationic amplitudes fall while the oxygen amplitude rises to a maximum near~$500~\si{\per\centi\meter}$. A transition from cation- to oxygen-dominated motion occurs near~$300~\si{\per\centi\meter}$, where the amplitudes cross. Note that the Zn (oxygen) amplitude reaches a global maximum (minimum) near~$150~\si{\per\centi\meter}$. It is reasonable to conclude that the smaller peak in the optical conductivity (Fig.~\ref{fig:optical-conductivity}) is caused by low-frequency oscillations of~Zn ions against a stationary oxygen sublattice.

\subsubsection{\label{sss:phase-quotient}Phase quotient}
The phase quotient introduced by Bell and Hibbins-Butler~\cite{bell_JPhysC_1975} generalizes the notions of acoustic/optical phonons to disordered media. It is defined by
\begin{equation}
    \phi(\omega_s)
    =\frac{\sum_{\langle bb'\rangle}\sum_\alpha\varepsilon_\alpha(b;s)\,\varepsilon_\alpha(b';s)}
    {\sum_{\langle bb'\rangle}\abs{\sum_\alpha\varepsilon_\alpha(b;s)\,\varepsilon_\alpha(b';s)}}
    \label{eq:PQ}
\end{equation}
where the summation~$\sum_{\langle bb'\rangle}$ extends over all nearest-neighbours~$\{b'\}$ of~$b$. If $\phi(\omega_s)=+1$ nearest-neighbours vibrate in-phase and the vibrations are acoustic-like; if $\phi(\omega_s)=-1$ nearest-neighbours vibrate out-of-phase and the vibrations are optical-like.

\begin{figure}[htbp]
    \centering
    \includegraphics{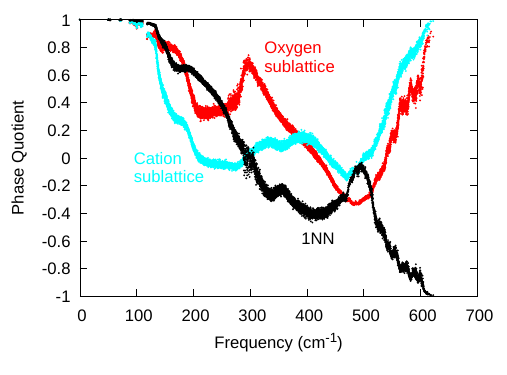}
    \caption{Phase quotient of HEO. The curves marked `sublattice' are calculated within a species-sublattice, \ie~with the summation over $\{b'\}$ in Eq.~\ref{eq:PQ} restricted to nearest-neighbours within the corresponding sublattice.}
    \label{fig:phase-quotient}
\end{figure}

The phase quotient of HEO is shown by the series marked `1NN' in Fig.~\ref{fig:phase-quotient}. As in ordered solids, the vibrations are highly acoustic-like at low frequency, near the Goldstone modes. As frequency increases, a transition to optical-like modes occurs. The peak near $500~\si{\per\centi\meter}$ coincides with the maximum (minima) of the oxygen (cation) relative amplitudes in Fig.~\ref{fig:relative-amplitude}. Since Fig.~\ref{fig:relative-amplitude} implies that it is only oxygen that moves at $500~\si{\per\centi\meter}$, the phase quotient is close to zero there, `midway' between acoustic and optical vibrations.

Figure~\ref{fig:phase-quotient} also presents modified phase quotients with the summation over $\{b'\}$ restricted to nearest-neighbours in the cation/oxygen sublattices separately. Interestingly, the phase quotient within the oxygen sublattice peaks at $300~\si{\per\centi\meter}$ where we see a transition from cation- to oxygen-dominated motion.

\subsubsection{\label{sss:polarization}Polarization}
\begin{figure*}[htbp]
    \centering
    \includegraphics{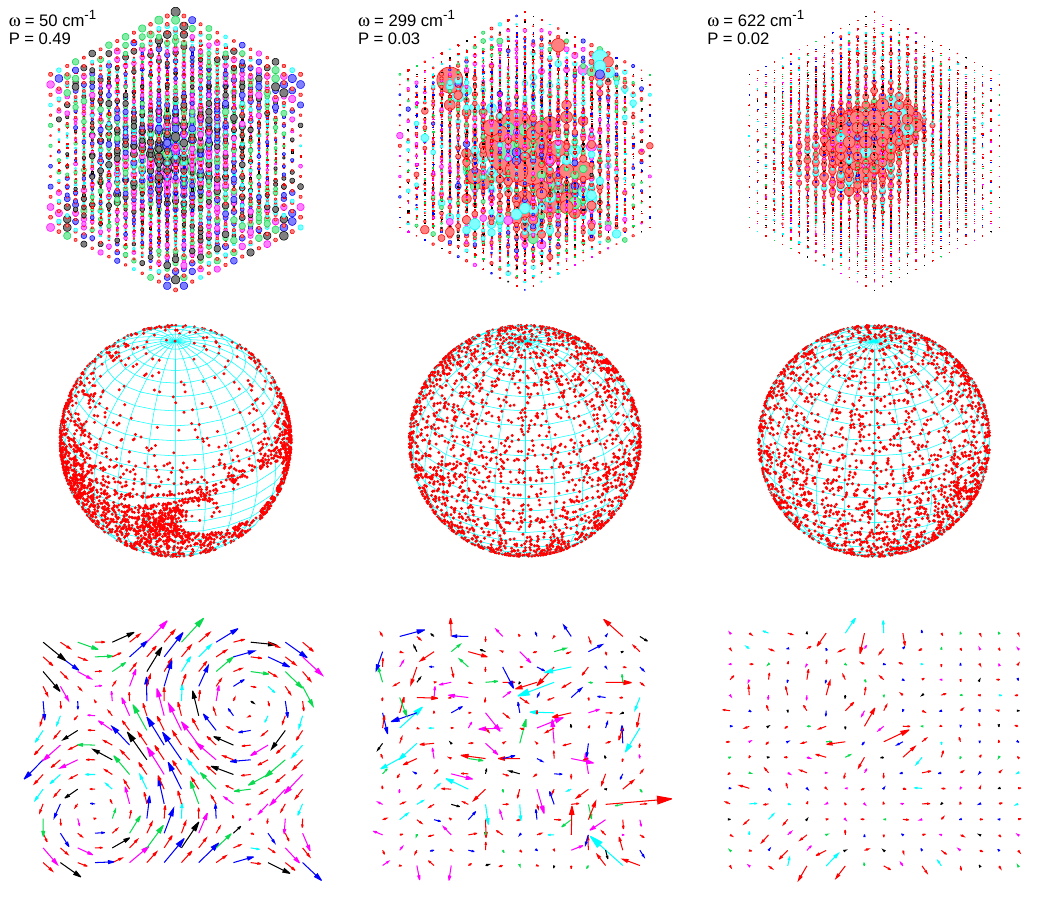}
    \caption{(Top row): Spatial distribution of vibrational energy. A larger bubble indicates an atom which is vibrating more intensely. The atomic color scheme is defined in Fig.~\ref{fig:relative-amplitude}. (Middle row): Polarizations projected onto the unit sphere. (Bottom row): Polarizations in the $z=\frac12$ plane. Mode frequencies and participation ratios (see Sect.~\ref{sss:participation-ratio}) are listed in the upper left.}
    \label{fig:3-3}
\end{figure*}

As discussed in Sec.~\ref{sec:intro}, nearest-neighbour eigenvector components are correlated in a propagon and uncorrelated in a diffuson/locon. To investigate the nature of nearest-neighbour vibrations in HEO, we project eigenvectors onto the unit sphere~\cite{yao_JACS_2022} according to
\begin{equation}
    \varepsilon_\alpha(b;s)
    \rightarrow
    \frac{\varepsilon_\alpha(b;s)}{\sum_\alpha\abs{\varepsilon_\alpha(b;s)}^2}.
    \label{eq:projections}
\end{equation}
Each dot in Fig.~\ref{fig:3-3} (middle row) represents the arrowhead of a vector pointing in the direction along which a given atom oscillates. Observe that the polarizations in the leftmost column are not uniformly distributed over the unit sphere. This suggests correlated motion across the system, as expected in a propagon mode. In contrast,  the $299~\si{\per\centi\meter}$ and $622~\si{\per\centi\meter}$ modes are incoherent with seemingly random distributions. This is consistent with the modes being diffusons or locons.

The polarizations in the $z=\frac12$ plane are shown explicitly in the bottom row of Fig.~\ref{fig:3-3}. The $z$-component of each $\varepsilon_\alpha(b;s)$ has been suppressed for ease of visualization. Vortex-like structures appear in the propagon mode, indicating plane-wave-like character. The orientations in the $299~\si{\per\centi\meter}$ and $622~\si{\per\centi\meter}$ modes appear to be random. 

\subsubsection{\label{sss:vortex-density}Vortex density}
The arrow plots of Fig.~\ref{fig:3-3} show a few `vortex' features in the low-energy modes. We argue that this can serve as a signature for plane-wave-like coherence. In a plane wave, the atomic-displacement-field varies in a smooth manner, akin to a flow-field. This smooth nature manifests as low vorticity. In contrast, high-energy modes have displacements that vary drastically at the atomic scale. This leads to large vorticity --  a large number of plaquettes (the squares formed by nearest-neighbour cation-oxygen bonds) have displacement fields that wind by $\pm 2\pi$. To formalize this argument, we define the `vortex density' as  
\begin{equation}
    \rho_v(s)
    =\frac{1}{N_p}\sum_{\ell}\delta_{\theta_\ell,2\pi}
    \label{eq:vortex-density}
\end{equation}
where $\sum_\ell$ runs over the centers of the $N_p=256$ plaquettes in the $z=\frac12$ plane and $\theta_\ell$ is the winding angle obtained by traversing the $\ell$th plaquette counterclockwise~\cite{alvarez_PRB_2001}. The Kronecker delta picks out only the `up'-vortices; \ie~those vortices with curl directed out-of-the-page. It follows from Stokes' theorem that up- and down-vortices are created in pairs~\cite{gerber_IOP_2015}.

\begin{figure}[htbp]
    \centering
    \includegraphics{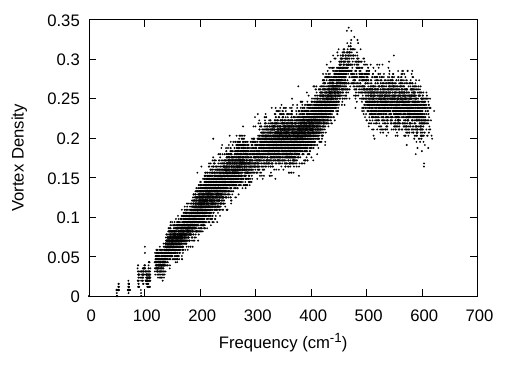}
    \caption{Vortex density of polarizations in the $z=\frac12$ plane.}
    \label{fig:vortex-density}
\end{figure}

The vortex density of Eq.~\ref{eq:vortex-density} is plotted in Fig.~\ref{fig:vortex-density} as a function of frequency. To interpret this result, we draw an analogy to the Berezinskii-Kosterlitz-Thouless (BKT) phase transition~\cite{gerber_IOP_2015}. In an XY ferromagnet, low-temperature configurations are smooth with no vortices. In contrast, high-temperature configurations appear to be random with high vorticity. Here, we tune the energy of eigenmodes rather than temperature. At low energies, vorticity is non-zero, but small -- indicating smoothness of the displacement field. As energy increases, vorticity grows indicating loss of coherence. 

We propose that the eigenmodes transition from propagon- to diffuson-character via a crossover. The former are smooth with low vortex density while the latter have large vortex densities. We propose a heuristic criterion to locate the crossover: $\rho_v \sim 10 \%$. As seen from Fig.~\ref{fig:vortex-density}, this criterion places the propagon-diffuson crossover at $\sim 200~\si{\per\centi\meter}$.

\subsubsection{\label{sss:spectral-weight}Spectral weight}
It is worthwhile to investigate the plane-wave character of the eigenmodes. Following Allen~\etal~\cite{allen_PhilMagB_1999}, we define the spectral weight associated with the mode $s$ as
\begin{equation}
    w_\alpha(q;s)
    =\abs{\sum_b\varepsilon_\alpha(b;s)\,e^{iq\cdot b}}^2
    \label{eq:spectral-weight}
\end{equation}
where $q_\alpha=2\pi n_\alpha/Na$ is $\alpha$th component of the wavevector, $a$ is the average lattice constant calculated from the relaxed cell lengths in Table~\ref{tab:crystal-structure}, $N=8$ is the number of times the conventional cubic unit cell was extruded along each Cartesian axis to generate the HEO supercell, and $n_\alpha=0,1,\dots,N_\alpha-1$ are non-negative integers.

In Fig.~\ref{eq:spectral-weight} we plot $\sum_\alpha\abs{w_\alpha(q;s)}^2$ for each of the allowed $\{q\}$ in three select eigenmodes. Only the $q=0$ term contributes to Eq.~\ref{eq:spectral-weight} in the Goldstone mode. The propagon mode is a superposition of several plane waves with small $q$, while the diffuson modes contains contributions from many different~$\{q\}$.
\begin{figure*}[htbp]
    \centering
    \includegraphics{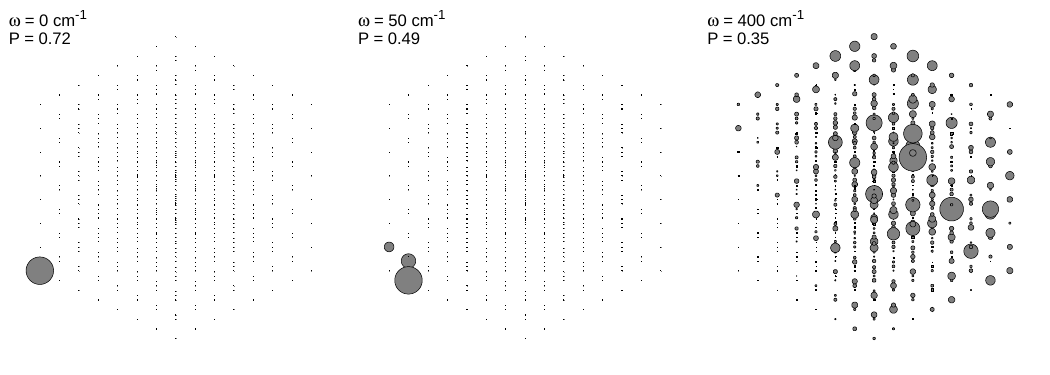}
    \caption{Spectral weight on each atom in select eigenmodes. In all cases, the diameter of the largest bubble has been normalized to unity for ease of visualization.}
    \label{fig:spectral-weight}
\end{figure*}

\subsection{\label{ssec:localized-modes}Localized vibrational modes}
We next explore the possibility of localization in HEO by examining various diagnostics.

\subsubsection{\label{sss:participation-ratio}Participation ratio}
The participation ratio $P$ defined by
\begin{equation}
    P(\omega_s)
    =\frac{1}{N}
    \times
    \frac{\left[\sum_{b}\sum_\alpha\abs{\varepsilon_{\alpha}(b;s)}^2\right]^2}
    {\sum_{b}\left[\sum_\alpha\abs{\varepsilon_{\alpha}(b;s)}^2\right]^2}
    \label{eq:PR}
\end{equation}
is an order parameter for Anderson localization in disordered systems. Here $N$ is the number of atoms in the supercell. If in some mode, the vibration is localized on a single atom, then $P=N^{-1}$. Conversely, if all atoms contribute with equal weight to a mode, then $P=1$. In practice, $P<0.1$ is often used as the cutoff for localized modes~\cite{seyf_JAP_2016}. We adopt a more conservative criterion for localization:~$P<0.05$.

\begin{figure}[htbp]
    \centering
    \includegraphics{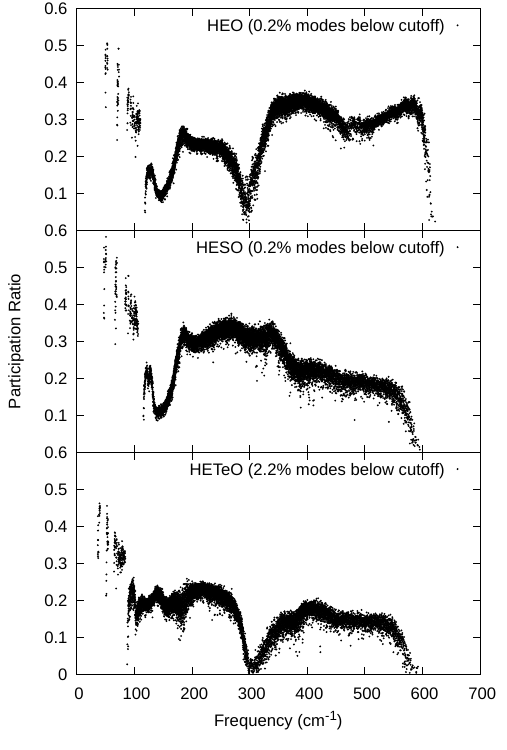}
    \caption{Participation ratio of HEO, HESO, and HETeO.}
    \label{fig:participation-ratio}
\end{figure}

Figure~\ref{fig:participation-ratio} compares the participation ratios of single 4096-atom disorder realizations of HEO, HESO, and HETeO. Each system has a mobility edge near~$600~\si{\per\centi\meter}$ beyond which $P$ falls rapidly towards zero. 

\begin{figure}[htbp]
    \centering
    \includegraphics{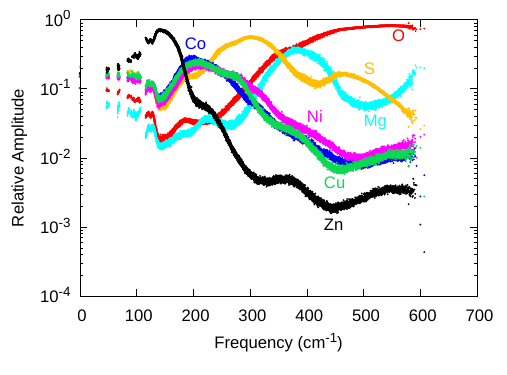}
    \caption{Relative amplitudes in HESO.}
    \label{fig:HESO-RA}
\end{figure}

We find modes with $P<0.05$ in two regions: in the middle of the spectrum at around $300~\si{\per\centi\meter}$ and at the top end near $600~\si{\per\centi\meter}$. 
The low-$P$ modes near $300~\si{\per\centi\meter}$ may originate from the transition from cation- to oxygen-dominated motion seen in the relative amplitudes (Fig.~\ref{fig:relative-amplitude}). Further support for this idea comes by noting that the relative amplitude of sulfur ions in HESO (Fig.~\ref{fig:HESO-RA}) reaches a global maximum at~$300~\si{\per\centi\meter}$. Hence there is no well-defined transition from cation- to oxygen-dominated motion in HESO, and hence no dip in its participation ratio. 

In HETeO, the $300~\si{\per\centi\meter}$ minimum is considerably wider and contains modes with very low $(P\sim10^{-3})$ participation ratios. This phenomenon of localized mid-spectrum modes can be understood with reference to the parent oxide NiO. It has two phonon bands, separated by a band gap~\cite{reichardt_JPhysC_1975}. In broad terms, the lower and upper bands have acoustic and optical character, respectively. Upon introducing disorder, we may view each band as developing an independent mobility edge above which excitations are localized. In this point of view, the localized modes around $300~\si{\per\centi\meter}$ are `acoustic locons' while those near $600~\si{\per\centi\meter}$ are `optical locons'.
\begin{figure}[htbp]
    \centering
    \includegraphics{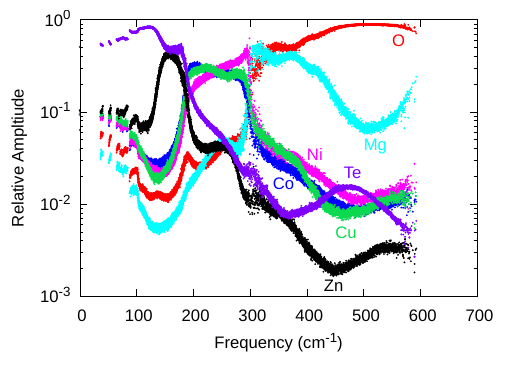}
    \caption{Relative amplitudes in HETeO.}
    \label{fig:HETeO-RA}
\end{figure}

\subsubsection{\label{ss:bubble}Spatial distribution of vibrational energy}
In the top row of Fig.~\ref{fig:3-3}, each ion in a 4096-atom disorder realization of HEO is represented by a sphere with diameter proportional to $\sum_\alpha\abs{\varepsilon_\alpha(b;s)}^2$. Thus a larger sphere indicates an ion which is vibrating more intensely, and vice-versa. Periodic boundary conditions have been invoked to shift the atom~$b$ with the largest $\sum_\alpha\abs{\varepsilon_\alpha(b;s)}^2$ to the center of the supercell. Visual inspection suggests that the $50~\si{\per\centi\meter}$ and $299~\si{\per\centi\meter}$ modes are extended, while the $622~\si{\per\centi\meter}$ mode is localized.

\begin{figure}[htbp]
    \centering
    \includegraphics{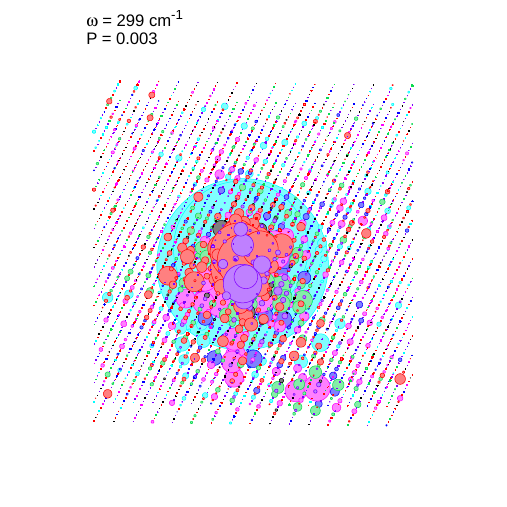}
    \caption{Spatial distribution of vibrational energy in a $299~\si{\per\centi\meter}$ mode in HETeO. The atomic color scheme is defined in Fig.~\ref{fig:HETeO-RA}.}
    \label{fig:HETeO-bubble}
\end{figure}

Figure~\ref{fig:HETeO-bubble} shows the distribution of vibrational energy in a $299~\si{\per\centi\meter}$ mode in HETeO. It appears that this mode is considerably more localized than the $299~\si{\per\centi\meter}$ mode in HEO.

\subsubsection{\label{sss:local-atomic-environment}Local atomic environment}
\begin{figure}[htbp]
    \centering
    \includegraphics{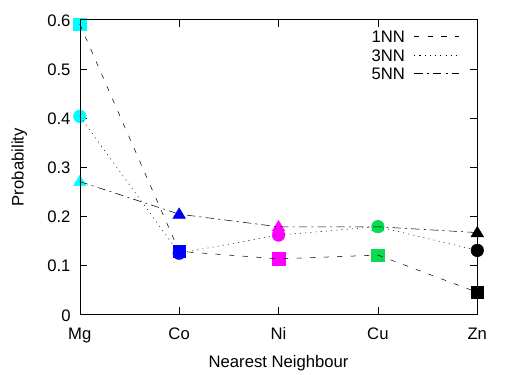}
    \caption{Nearest-neighbours of the oxygen ion with the largest eigenvector component in the high-frequency mode. The dashed lines are guides to the eye.}
    \label{fig:atomic-environment}
\end{figure}

It is natural to ask whether or not short-range correlations influence the formation of the $620~\si{\per\centi\meter}$ locon in Fig.~\ref{fig:3-3}. We denote the oxygen ion with the largest eigenvector component in the highest-frequency mode as (O)$_{max}$. Figure~\ref{fig:atomic-environment} shows the distribution of $n$th nearest-neighbour cations surrounding (O)$_{max}$. The distribution was averaged over fifty 4096-atom disorder realizations. There is a strong preference for~Mg ions as nearest-neighbours. By the fifth nearest-neighbour shell ($\sqrt{5}a/2\sim4.7~\si{\angstrom}$ away from (O)$_{max}$), the distribution is practically uniform. Physically, the locon is situated in a cage of lighter Mg ions. There must then be a surplus of Zn and heavier cations at distances from (O)$_{max}$ greater than the fifth nearest-neighbour shell, which insulate the vibration from the rest of the supercell. The nearest-neighbour bond lengths surrounding (O)$_{max}$ were also investigated, but were equal within error to the bulk results in Table~\ref{tab:bond-lengths}.

\subsubsection{\label{sss:localization-length}Localization length}
Allen~\etal~\cite{allen_PhilMagB_1999} observed that the eigenvector components of a locon in $\alpha$-Si decay according to
\begin{equation}
    \abs{\varepsilon(b;s)}
    \propto
    \exp(-\frac{\abs{b-b_{max}}}{\xi_s})
    \label{eq:loclen}
\end{equation}
where $b_{max}$ is the atom with the largest eigenvector component and $\xi_s$ is the localization length of mode $s$, which should be much smaller than the size of the supercell.
\begin{figure}[htbp]
    \centering
    \includegraphics{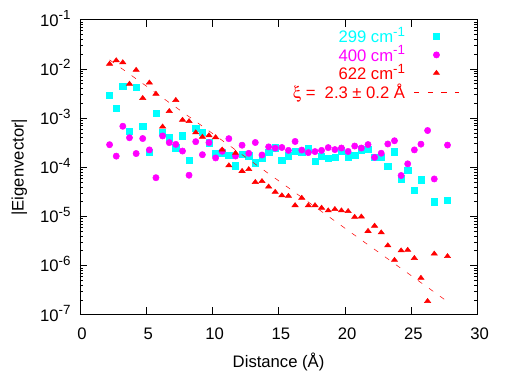}
    \caption{Spatial decay of vibrational eigenvector components. The $622~\si{\per\centi\meter}$ mode is localized within $\xi=2.4(3)~\si{\angstrom}$.}
    \label{fig:localization-length}
\end{figure}
The decay profiles of three modes in a 4096-atom disorder realization of HEO are shown in Figure~\ref{fig:localization-length}. The~$622~\si{\per\centi\meter}$ mode is well-described by Eq.~\ref{eq:loclen} for all space and is localized within ~$\xi=2.3(2)~\si{\angstrom}$. Neither the $299~\si{\per\centi\meter}$ nor the $400~\si{\per\centi\meter}$ modes agree well with Eq.~\ref{eq:loclen}. This is particularly surprising for the $299~\si{\per\centi\meter}$ mode, which should be a locon according to its low participation ratio. Allen~\etal~have observed similar decay profiles for $590~\si{\per\centi\meter}$ diffusons in $\alpha$-Si~\cite{allen_PhilMagB_1999}.

\begin{figure}[htbp]
    \centering
    \includegraphics{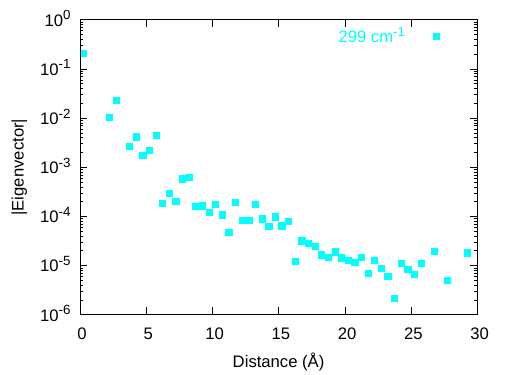}
    \caption{Spatial decay of vibrational eigenvector components in $299~\si{\per\centi\meter}$ mode in HETeO.}
    \label{fig:HETeO-loclen}
\end{figure}

Figure~\ref{fig:HETeO-loclen} shows the decay profile of the $299~\si{\per\centi\meter}$ mode in HETeO depicted in  Figure~\ref{fig:HETeO-bubble}. Visual inspection suggests Figure~\ref{fig:HETeO-loclen} may be well-described by a double-exponential fit; \ie~this mode is a superposition of two locons; one with very small~$\xi$ near $0~\si{\angstrom}$, and a second locon with larger~$\xi$ persisting over longer distances.

\subsubsection{\label{sss:level-statistics}Level statistics}
\begin{figure}[htbp]
    \centering
    \includegraphics{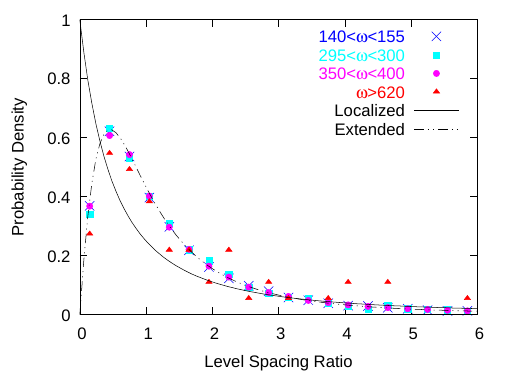}
    \caption{Distributions of level spacing ratios. All frequencies are in units of~$\si{\per\centi\meter}$. The extended and localized distributions refer to Eqs.~\ref{eq:wigner-dyson} and~\ref{eq:poisson}, respectively.}
    \label{fig:level-statistics}
\end{figure}

The level spacing $\epsilon_s$ between adjacent squared eigenfrequencies (\ie~eigenvalues of the dynamical matrix) is defined as $\epsilon_s=\omega_{s+1}^2-\omega_s^2$. The level spacing ratio $r_s$ is in turn defined as $r_s=\epsilon_s/\epsilon_{s-1}$. Random matrix theory predicts that the $\{r_s\}$ of an extended system in the Gaussian Orthogonal Ensemble are distributed according to
\begin{equation}
    P(r)=\frac{27}{8}\,\frac{r+r^2}{(1+r+r^2)^{5/2}},
    \label{eq:wigner-dyson}
\end{equation}
while for a localized system the distribution is~\cite{atas_PRL_2013}
\begin{equation}
    P(r)=\frac{1}{(1+r)^2}.
    \label{eq:poisson}
\end{equation}
Figure~\ref{fig:level-statistics} shows the distribution of level spacing ratios for several frequency windows in HEO. An average was performed over fifty 4096-atom disorder realizations. As expected, the diffusons near $147~\si{\per\centi\meter}$ and $400~\si{\per\centi\meter}$ are well-described by Eq.~\ref{eq:wigner-dyson}. The level statistics also suggest that modes near $300~\si{\per\centi\meter}$ are delocalized despite satisfying $P<0.05$. This is consistent with the analysis of the localization lengths in Sec.~\ref{sss:localization-length}. No range of frequencies could be found which resulted in an agreement between Eq.~\ref{eq:loclen} and modes beyond the mobility edge. This may be due to a paucity of data points: each $4096$-atom disorder realization contains only $\mathcal{O}(10^{0})$ high-frequency locons, compared to~$\mathcal{O}(10^{3})$ diffusons between 350--400$~\si{\per\centi\meter}$.

\subsubsection{\label{sss:correlation-amplitude}Correlation amplitude}
We propose that localized and extended modes can further be differentiated by monitoring the time evolution of a vibrational wavepacket. Consider a cubic supercell $C_1$ of HEO wrapped in the center of a larger HEO environment $\bar{C_1}$. The union of $C_1\cup\bar{C_1}$ defines a larger supercell $C_2$. If $C_1$ is not too small, $C_2$ should not look too dissimilar from periodic repetitions of $C_1$. Now, suppose all the atoms in $C_2$ have zero initial velocity at time $t=0$. Furthermore, let all the atoms in $C_1$ be displaced according to one of the eigenmodes of $C_1$, and let all the atoms $\bar{C_1}$ be fixed to zero initial displacement. This defines a wavepacket (consisting of the atoms in $C_1$) initially having potential energy only.

In analogy with quantum dynamics, it is possible to define, following Allen and Kelner~\cite{allen_AJP_1998}, a correlation amplitude $C(t)$ for a wavepacket of HEO.  This quantity represents the overlap of an initial local wavepacket (with deviations from equilibrium localized in a small region) with the evolved state at a later time. The details are left to the Appendix. The result is
\begin{equation}
    C(t)
    =(\Omega^T\varepsilon^\dag\vb{s})^T\,e^{-i\Omega t}\,(\Omega^T\varepsilon^\dag\vb{s})
    \label{eq:corr-amp}
\end{equation}
where $\Omega$ is a diagonal matrix of the $C_2$ eigenfrequencies, $\varepsilon$ is a unitary matrix of the $C_2$ eigenvectors, and $\vb{s}$ is a column vector with components equal to zero if an atom $b\in\bar{C_1}$, and equal to a component of a $C_1$ eigenvector if $b\in C_1$. If the wavepacket consisted of locons, the wavefront should not spread far and the state of $C_2$ at a later time $t>0$ should strongly resemble the initial state at time $t=0$; \ie~the ratio $\abs{C(t)/C(0)}$ of the correlation amplitudes should not deviate far from unity. Conversely, a wavepacket composed of propagons or diffusons should spread over $C_2$ and $\abs{C(t)/C(0)}$ should decay to zero.

\begin{figure}[htbp]
    \centering
    \includegraphics{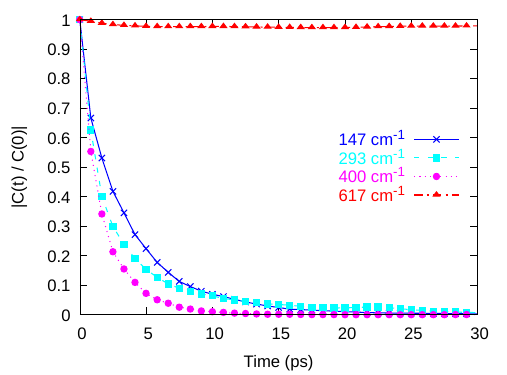}
    \caption{Correlation amplitudes of select modes in HEO.}
    \label{fig:correlation-amplitude}
\end{figure}

These predictions were tested by embedding a 1728-atom cluster of HEO in an 8000-atom environment. It may be seen in Figure~\ref{fig:correlation-amplitude} that the $617~\si{\per\centi\meter}$ mode (the highest-frequency locon in $C_1$)  remains very close to unity at all times, while the other modes decohere quickly.

\section{Summary and Discussion}
The experimental crystal structure, bond lengths, and optical conductivity of HEO are well-reproduced by modelling the ions as charged shells interacting through the Buckingham potential. We ascribe the weak mode in the optical conductivity discovered by Afsharvosoughi and Crandles to low-frequency, acoustic-like vibrations of Zn ions against a stationary oxygen sublattice. The combination of participation ratio, mode polarizations, localization length, and correlation amplitude suggest localized modes exist in HEO around $620~\si{\per\centi\meter}$, beyond a mobility edge. Modes in the vicinity of the transition from cation- to oxygen-dominated motion at~$300~\si{\per\centi\meter}$ have a tendecy to localize, as reflected in low values of the participation ratio. Indeed, they may be driven to localize by increasing the mass disorder in the oxygen sublattice, as in the proposed HETeO.

Future studies, with larger system sizes, may find sharper signatures of localization. For example, it may be possible to reproduce the Wigner-Dyson and Poisson statistics (Eqs.~\ref{eq:wigner-dyson} and~\ref{eq:poisson}) expected from theoretical considerations. This may require specialized numerical techniques such as the kernel polynomial method~\cite{weisse_PRE_2006} to handle large matrices. Future studies may also be able to distinguish between propagons and diffusons by quantifying the spatial spread of energy. Propagons are expected to spread with a ballistic front, while diffusons are expected to spread diffusively. 

Localized vibrational modes in HEO may be experimentally observable by monitoring the spatial dependence of IR transmission. According to the participation ratio, a $300~\si{\per\centi\meter}$ IR beam will excite diffusons. Sweeping the beam across a sample of HEO should not have a significant effect on the transmitted intensity. However, the transmitted intensity of $620~\si{\per\centi\meter}$ light should plummet whenever the beam impinges upon a locon. Such an experiment will require small spot sizes that are comparable to typical locon mode radii. Recent advances with MINFLUX nanoscopy~\cite{gwosch_nature_2020} have achieved spot sizes as small as 1-3~$\si{\nano\meter}$ -- a few times larger than the locon in Fig.~\ref{fig:localization-length}. 

We have shown that admixing tellurium with oxygen to form HETeO creates mid-frequency localized modes. This can motivate synthesis efforts to increase entropy in the anion sublattice as well. The effects of mass and force constant disorder can further be studied in medium-entropy oxides~\cite{crandles_JAP_2021}, where the configurational entropy is considerably reduced, or in other high-entropy oxides with different crystal structures. Finally, the localization of vibrations in HEO motivates future investigations into localization with other types of excitations. For example, neutron scattering studies have demonstrated that HEO develops antiferromagnetic order at $T\sim112~\si{\kelvin}$~\cite{zhang_ChemOfMat_2019}. The magnetic collective excitations share many similarities to the vibrational modes studied in this work. It is conceivable that high entropy oxides may host localized spin wave modes. We may find analogous localization with electronic states as well, \eg~in the related family of high entropy alloys~\cite{yeh_AEM_2004}.

\begin{acknowledgments}
This research was supported by the National Science and
Engineering Council of Canada (NSERC) and the Ontario Provincial Government (QEII-GSST). We would like to thank Dr.~Eric~de~Giuli (Toronto Metropolitan University), Dr.~Jiri~Hlinka (Czech Academy of Sciences), and Dr.~Edward~Sternin (Brock University) for their many useful comments and discussions.
\end{acknowledgments}

\appendix
\section{Wavepacket Dynamics}
The general, time-dependent solution $u_\alpha(\ell b;t)$ to the equations of motion in a crystal with $\mathcal{N}$ unit cells is~\cite{srivastava_1990}
\begin{align}
    u_\alpha(\ell b;t)
    &=\frac{1}{\sqrt{\mathcal{N}m_b}}\sum_{qs}Q(qs;t)\,\varepsilon_\alpha(b;qs)\,e^{iq\cdot\ell}
    \label{eq:general-soln}
\end{align}
where~$\ell$ labels the unit cells, $m_b$ is the mass of the $b$th atom, $q$ is one of the wavevetors allowed by periodic boundary conditions, and $\{Q\}$ are the normal coordinates defined by
\begin{equation}
    Q(\ell b;t)
    =Q(qs;0)\,\cos\omega_{qs}t+
    \frac{\dot{Q}(qs;0)}{\omega_{qs}}\,\sin\omega_{qs}t.
    \label{eq:normal-coords}
\end{equation}
By substituting Eq.~\ref{eq:normal-coords} into Eq.~\ref{eq:general-soln} and exploiting the completeness and orthonormality of the vibrational eigenvectors $\{\varepsilon_\alpha(b;qs)\}$, it is possible to express the initial $\{Q,\dot{Q}\}$ in terms of the initial $\{u,\dot{u}\}$:
\begin{align}
    \notag
    Q(qs;0)=\frac{1}{\sqrt{\mathcal{N}}}\sum_{b\alpha}\sqrt{m_b}\,\varepsilon_\alpha^*(b;qs)\\\times\sum_\ell u_\alpha(\ell b;0)\,e^{-iq\cdot\ell}.
    \label{eq:initial-Q}
\end{align}
An identical equation holds with~$Q$ replaced by~$\dot{Q}$ and~$u$ replaced by~$\dot{u}$. Consider now the application of the foregoing equations to a large supercell. To first approximation the Brillouin zone consists of the single point $q=0$; \ie~converged phonon properties are obtained from a $\Gamma$-point calculation only. From this it follows that $\ell=0$ and $\mathcal{N}=1$. Substituting Eqs.~\ref{eq:normal-coords} and~\ref{eq:initial-Q} into Eq.~\ref{eq:general-soln} produces
\begin{align}
    \notag
    u_\alpha(b;t)
    &=\frac{1}{\sqrt{m_b}}\sum_s\varepsilon_\alpha(b;s)\,\cos\omega_st\\&\times\sum_{b'\alpha'}\sqrt{m_{b'}}\,\varepsilon_{\alpha'}^*(b';s)\,u_{\alpha'}(b';0)
    \label{eq:supercell-general-soln}
\end{align}
where the labels $\ell=q=0$ have been discarded. The Dirac notation used by Allen and Kelner~\cite{allen_AJP_1998} is now introduced through the correspondence rules
\begin{align*}
    u_\alpha(b;t)
    &\leftrightarrow\ip{b\alpha}{u(t)}\\
    \varepsilon_\alpha(b;s)
    &\leftrightarrow\ip{b\alpha}{s}\\
    \varepsilon_\alpha^*(b;s)
    &\leftrightarrow\ip{s}{b\alpha}\\
    \cos\omega_st\,\delta_{ss'}
    &\leftrightarrow\mel{s}{\hat{C}(t)}{s'}\\
    m_b\,\delta_{bb'}\delta_{\alpha\alpha'}
    &\leftrightarrow\mel{b\alpha}{\hat{M}}{b'\alpha'}
\end{align*}
Here $\{m_b\}$ are the eigenvalues of the mass operator $\hat{M}$ which is diagonal in the direct space basis $\{\ket{b\alpha}\}$. Similarly, $\{\cos\omega_st\}$ are the eigenvalues of the operator $\hat{C}(t)$ which is diagonal in the eigenbasis $\{\ket{s}\}$. It is easily shown that Eq.~\ref{eq:supercell-general-soln} can be recast as
\begin{widetext}
\begin{equation}
    \ip{b\alpha}{u(t)}
    =\sum_{ss'}\sum_{b'\alpha'}\sum_{b''\alpha''}\sum_{b'''\alpha'''}\mel{b\alpha}{\hat{M}^{-\frac12}}{b'\alpha'}\ip{b'\alpha'}{s}\mel{s}{\hat{C}(t)}{s'}\ip{s'}{b''\alpha''}\mel{b''\alpha''}{\hat{M}^{\frac12}}{b'''\alpha'''}\ip{b'''\alpha'''}{u(0)}.
 \label{eq:wideeq}
\end{equation}
\end{widetext}
Summations $\sum_s\dyad{s}{s}$ or $\sum_{b\alpha}\dyad{b\alpha}{b\alpha}$ over dyads generate resolutions of unity by completeness. Hence Eq.~\ref{eq:wideeq} can be succinctly written as
\begin{equation}
    \ket{u(t)}=\hat{M}^{-\frac12}\hat{C(t)}\hat{M}^{\frac12}\ket{u(0)}
    \label{eq:positions-dirac}
\end{equation}
where we have used the fact that Eq.~\ref{eq:supercell-general-soln} holds for an arbitrary bra vector $\bra{b\alpha}$. Similarly, for the velocities
\begin{equation}
    \ket{v(t)}=-\hat{M}^{-\frac12}\hat{\Omega}\hat{S(t)}\hat{M}^{\frac12}\ket{u(0)}
    \label{eq:velocities-dirac}
\end{equation}
where $\hat{\Omega}$ and $\hat{S}(t)$ are diagonal in the eigenbasis with eigenvalues $\{\omega_s\}$ and $\{\sin\omega_st\}$, respectively. Following Allen and Kelner, consider now the vector
\begin{align}
    \notag
    \ket{W(t)}
    &=\hat{M}^{\frac12}\ket{v(t)}-i\hat{\Omega}\hat{M}^{\frac12}\ket{u(t)}\\
    &=e^{-i\hat{\Omega t}}\ket{W(0)}
    \label{eq:W(t)}
\end{align}
where used has been made of Eqs.~\ref{eq:positions-dirac} and~\ref{eq:velocities-dirac}. Since Eq.~\ref{eq:W(t)} has the same form as the Schr\"{o}dinger equation for state kets, with $e^{-i\hat{\Omega}t}$ playing the role of the time evolution operator, it follows that $\ket{W(t)}$ is the classical analogue of the quantum state of the system. Taking the overlap of $\ket{W(t)}$ with $\ket{W(0)}$ gives the formula for the correlation amplitude shown in Eq.~\ref{eq:corr-amp}.

\bibliography{apssamp}

\end{document}